\documentclass[preprint2]{aastex62}

\lefthead{Wada}
\righthead{Origin of NLR in Circinus}

\newcommand{\bea}{\begin{eqnarray} }
\newcommand{\eea}{\end{eqnarray}}

\newcommand{\oiii}{[O{\scriptsize III}] }

\newcommand{\oi}{[O{\scriptsize I}] }
\newcommand{\nii}{[N{\scriptsize II}] }
\newcommand{\sii}{[S{\scriptsize II}] }

\newcommand{\niiha}{[N{\scriptsize II}]/H$\alpha$ }
\newcommand{\siiha}{[S{\scriptsize II}]/H$\alpha$ }
\newcommand{\oiha}{[O{\scriptsize I}]/H$\alpha$ }
\newcommand{\oiiihb}{[O{\scriptsize III}]/H$\beta$ }
\begin{document}


\title{Circumnuclear Multi-phase Gas in Circinus Galaxy III:  Structure of the Nuclear Ionized Gas}
 
\author{%
Keiichi Wada
}
\affiliation{Kagoshima University, Graduate School of Science and Engineering, Kagoshima 890-0065, Japan}
\affiliation{Ehime University, Research Center for Space and Cosmic Evolution, Matsuyama 790-8577, Japan}
\affiliation{Hokkaido University, Faculty of Science, Sapporo 060-0810, Japan}
\correspondingauthor{Keiichi Wada}
\email{wada@astrophysics.jp}

\author{
Kensuke Yonekura
}%
\affiliation{Kagoshima University, Graduate School of Science and Engineering, Kagoshima 890-0065, Japan}
\affiliation{NEC Aerospace Systems, Ltd.}

\author{Tohru Nagao}%
\affiliation{Ehime University, Research Center for Space and Cosmic Evolution, Matsuyama 790-8577, Japan}


%


\begin{abstract}
{
We investigate the properties of the ionized gas irradiated by an active galactic nucleus (AGN)
based on our ``radiation-driven fountain" model for the nearest type-2 Seyfert galaxy, the Circinus galaxy \citep{wada2016}.
We conducted ``quasi-three dimensional" spectral analysis using the spectral synthesis code  C{\scriptsize LOUDY}
and obtained the surface brightness distributions of lines, such as H$\alpha$, H$\beta$, \oiii, \nii,  and \sii 
for the central 16-parsec region.
The ionized regions observed based on these lines show a conical morphology around the rotation axis, even if
we do not phenomenologically postulate the presence of an optically thick ``torus".  
This region also shows non-uniform internal structures, reflecting the inhomogeneous structure of 
fountain flows. 
Using ionization diagnostic diagrams, we investigated the spectral properties of the 
ionized gas.  The diagrams based on the line ratios of \niiha and \siiha show that most 
regions of the cone have the same properties as those in the narrow line regions (NLRs) in AGNs,
whereas using \oiha,  the central 10-pc regions
are rather LINER-like. 
 The gas density, temperature, and ionizing parameter 
in regions identified as ``NLR" are typically $n \sim 300-1500$ cm$^{-3}$,
$T \sim 1-3\times 10^4 $ K, and $ U \sim 0.01$, respectively.
The morphology and \oiii intensity are similar to the base of the observed \oiii cone in
the Circinus galaxy, implying some physical connections with the origin of the $\sim100$ parsec scale NLR.
}
\end{abstract}

\keywords{galaxies: active --  galaxies: nuclei -- galaxies: ISM -- radiative transfer}

\section{INTRODUCTION}
{
  In the standard picture of active galactic nuclei (AGNs),  
an AGN hypothetically consists of an accretion disk around a supermassive black hole, a broad emission line region, 
a dusty ``torus", and a narrow emission line region (NLR).
NLRs are spectroscopically characterized by narrow (several 100 km s$^{-1}$) emission lines, 
and they spatially extend from tens of parsecs to a few kiloparsecs
from the nucleus. Therefore, these regions can be spatially resolved in nearby AGNs,
which are characterized by conical structures with clumpy internal morphologies \citep[e.g.][]{marconi1994, schmitt1996, veilleux2001,sharp2010, muller2011}.
It has been determined that NLRs are gases photo-ionized by a power-law spectrum from the nucleus
\citep{davidson1979, evans1986, binette1996,komossa1997, nagao2006}, 
showing characteristic lines, such as 
H$\alpha$, H$\beta$, \oiii$\lambda$5007, \nii$\lambda$6583, and \sii$\lambda\lambda$6716,6731\footnote{
Shock excitation should be also considered to 
explain some lines, such as [FeII] and [PII], that
form NLRs in some Seyfert galaxies \citep{mouri2000, oliva2001, storchibergmann2009, terao2016}. See also \citet{kraemer2000b} for modeling the origin of 
emission lines in NGC 1068.}.}

{
The structures of the ISM in NLRs on the parsec scale are still not resolved, even in nearby AGNs. 
Various theoretical models of the emitting gas have been proposed,
in which the ISM is assumed to consist of single- or multi-component clouds illuminated by
nuclear radiation \citep[e.g.]{baldwin1995, ferguson1997, murayama1998, nagao2003, groves2004}
(See also a review by \citet{groves2007}).
Imaging observations and long-slit spectra acquired by the Hubble Space Telescope/WFPC2 and STIS have been used to 
study the fundamental three-dimensional geometry and conditions of the NLR gases in some nearby AGNs, such as NGC 1068, NGC 4151, Mrk 3, and Mrk 573
\citep[e.g.][]{kraemer2000a, crenshaw2000, das2005, das2006, das2007, fischer2010, crenshaw2010}.
These studies suggest that the gases in NLRs form
intrinsically biconical outflows consisting of multiple components, whose conditions are 
determined by the central radiation propagating through the media around AGNs.
}

%



{
However, the origin of such outflowing, multi-component gas is not clear, 
and the reason why the conical and clumpy morphology of NLRs is formed is still an open question.
The conical shape suggests that the radiation from the central source is 
spatially limited by the optically thick ``torus", that the distribution of the outflowing gas is intrinsically conical, or both.

Instead of assuming phenomenological models, in which the geometry and properties of the ``torus" and outflowing gas are
postulated, here we use a physics-motivated, three-dimensional hydrodynamic model.
Recently, we proposed a novel, dynamic picture of the ISM in the central tens of parsecs of AGNs, based on 
 three-dimensional radiation-hydrodynamic calculations, i.e., a radiation-driven fountain \citep{wada2012}.
 In this picture, the outflowing, multi-phase gas with dust is naturally formed, and we found that
it is not necessary to postulate a donut-like ``torus" to explain 
type-1 and -2 dichotomies in the spectral energy distribution (SED) \citep{schartmann2014} . 

In this paper, we focus on the spectral properties of the fountain flows illuminated by the AGN, to determine whether
they show the properties of the observed NLRs.
One should note that our current hydrodynamic model is still spatially limited, i.e. $r \leq16$ pc, therefore 
it may not be used to explain the general morphology of NLRs extended to several hundred parsecs; 
however, it can be compared with the central part of the NLR in the nearest ($D=$ 4.2 Mpc, \citet{tully2009}) type-2 Seyfert galaxy, the Circinus galaxy.
}



The structure of this paper is as follows. In \S 2, we briefly describe the input hydrodynamic model, i.e., the radiation-driven fountain \citep{wada2016}, 
and the model set-up for {\it quasi-multi-dimensional} C{\scriptsize LOUDY} simulations. Numerical results are shown in \S 3, and
their implications are discussed with respect to the observed NLR in \S 4. A summary is given in \S 5.

%
\section{CLOUDY SIMULATIONS BASED ON THE HYDRODYNAMIC MODEL}
%

\subsection{Input model: Radiation-driven fountain}
\citet{wada2012} proposed that the obscuring structures around 
AGNs, in which outflowing and inflowing gases are driven by 
radiation from the accretion disk, form 
a geometrically thick disk-like structure on the scale of a few parsecs to tens of parsecs.
The quasi-steady ``poloidal" circulation of gas, i.e., the ``radiation-driven fountain," may
 obscure the central source \citep{wada2015}. 
 The differences in SEDs 
 of typical type-1 and -2 Seyfert galaxies
 are reasonably explained by changing the viewing angle without assuming 
 a donut-like optically thick torus \citep{schartmann2014}.
\citet{wada2016} applied this radiation-driven fountain model with
an X-ray-dominated region chemistry to the central 16-pc region of
the Circinus galaxy.
They found that dense molecular gases {($n_{{\rm H}_2} \gtrsim 10^3$ cm$^{-3}$)} are mostly concentrated around
the equatorial plane, and {atomic gas (e.g., H$^0$ and C$^0$)} extends with a larger scale height.
There is also ``polar" emission in the mid-infrared band (12 $\mu {\rm m}$), which is associated with
bipolar outflows, as suggested in recent interferometric observations of nearby AGNs
 \citep[][]{hoenig2013,tristram2014}.
The viewing angle $\theta_v$ toward the nucleus should be larger than 75$^\circ$ (i.e., close to edge-on) to
explain the observed SED and 10 $\mu m$ absorption feature of the Circinus galaxy \citep{prieto2010}.

This best-fit model for the Circinus galaxy was used in \citet{wada2018}, and the results explain the structures of the cold molecular gas located on the outskirts of
the bright infrared region.
These results were derived from 3-D non-local thermodynamic equilibrium line transfer calculations
for $^{12}$CO lines. 
These results are consistent with various aspects of ALMA Cycle-4 observations,
as discussed in \citet{izumi2018}.

In this paper, we focus on a cone structure occupied by an inhomogeneous, diffuse, ionized gas
in this fountain model \citep{wada2016}. 
The outflows surrounded by geometrically thick atomic/molecular gas
could be the origin of the NLRs. 
To observe the spectral properties of the outflowing gas in the numerical model, 
we apply a spectral synthesis code, C{\scriptsize LOUDY} \citep{ferland2017}, to 
the same snapshot data used in \citet{wada2018}.

\subsection{Radiative transfer using C{\scriptsize LOUDY}}

As shown in Fig. 1, one eighth of the 3-D grid data (i.e., density and temperature) from the 256$^3$ Cartesian grid cells in the radiation-hydrodynamic simulation \citep{wada2018}
is reformed to uniformly spaced polar grid cells with $ (N_r, N_\theta, N_\phi) = (16, 16, 16)$\footnote{The three-dimensional density field of the radiation-driven fountain is not perfectly 
axisymmetric and is not symmetric in the equatorial plane, 
but we think that 1/8 of the total box (i.e., the volume in $x>0$, $y>0$, and $z>0$) is 
enough to study the emission line properties in fountain flows}. 
The radial size of each grid cell is 1 parsec.

We then ran the spectral synthesis code C{\scriptsize LOUDY} (version 17.02) \citep{ferland2017} for polar grid cells along the {\it radial direction}
from the central source to the outer boundary at $r=16$ pc (Fig. 1a).  
Therefore, C{\scriptsize LOUDY} is called 16$^3$ times to obtain the intensity distributions of the emission lines.
Note that the present calculation is {\it not fully} three-dimensional, in the sense that 
non-radial photons that emerge from other grid cells are not taken into account 
as the incident radiation for a C{\scriptsize LOUDY} call.

The SED of the central source is given using C{\scriptsize LOUDY}'s  \verb+AGN+ command, which assumes the spectral:
\[
f_\nu = \nu^{\alpha_{uv}} \exp(-h\nu/kT_{BB}) \exp(-kT_{IR}/h\nu) +a \nu^{\alpha_x},
\]
where $\alpha_{uv} = -0.5$, $T_{BB} = 10^5$ K, $\alpha_{x} = -1$, $a$ is a constant that yields $\alpha_{\rm ox} = -1.4$, and {$k T_{IR} = 0.01$ Ryd}.
The bolometric luminosity of the central source is $L_{bol} = 5 \times 10^{43}$ erg s$^{-1}$, and 
the flux changes depending on direction as $\cos \theta$ ($\theta $ is the angle from the z-axis),
as we assume that the UV radiation is dominated by the thin accretion disk.
We assume solar metallicity.  The filling factor is set to unity.
{One should note that the choice of the filling factor of unity does not mean that the outflowing gas 
is ``uniform". The radiation-driven fountain flows are
not uniform on the scale of several parsecs. However, we here do not assume unresolved clumpy internal structures in one grid cell for 
each C{\scriptsize LOUDY} calculation. For simplicity, we intend to assume the least number of free parameters.}

The transmitted SED, calculated with C{\scriptsize LOUDY},
is used as an incident SED for the next outward radial cell,
and this procedure is repeated up to the outer edge (i.e., $r=16$ pc) for a given radial ray.
The line intensities, such as those of 
H$\alpha$, H$\beta$, \oiii$\lambda$5007, \oi$\lambda$6300, \nii$\lambda$6583, and \sii$\lambda\lambda$6716,6731,
are collected using the ``\verb+save line list+" command and are integrated along the line-of-sight ($y$-direction) to obtain surface brightness maps.

\begin{figure}[h]
\centering
\includegraphics[width = 7cm]{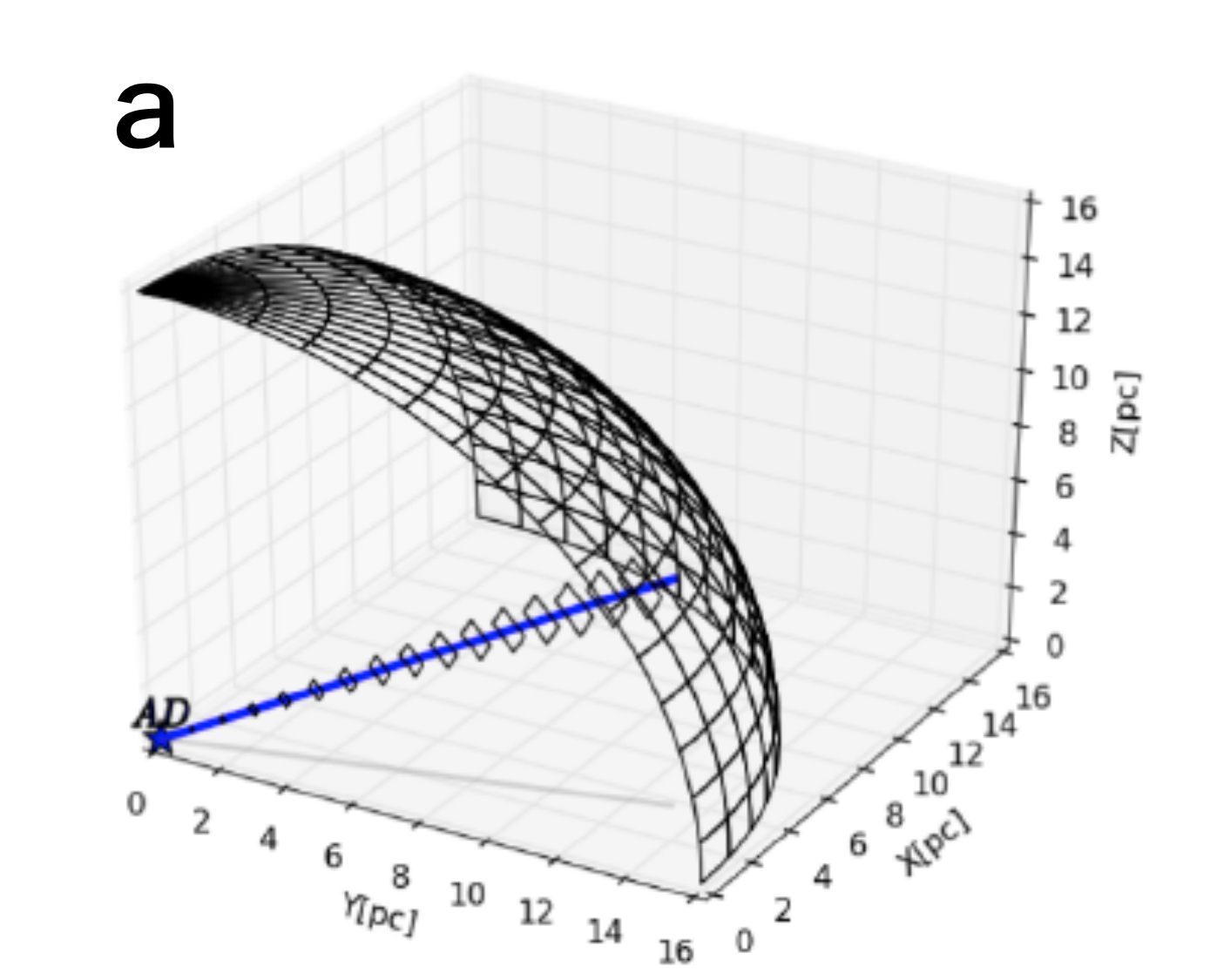}
\includegraphics[width = 6cm]{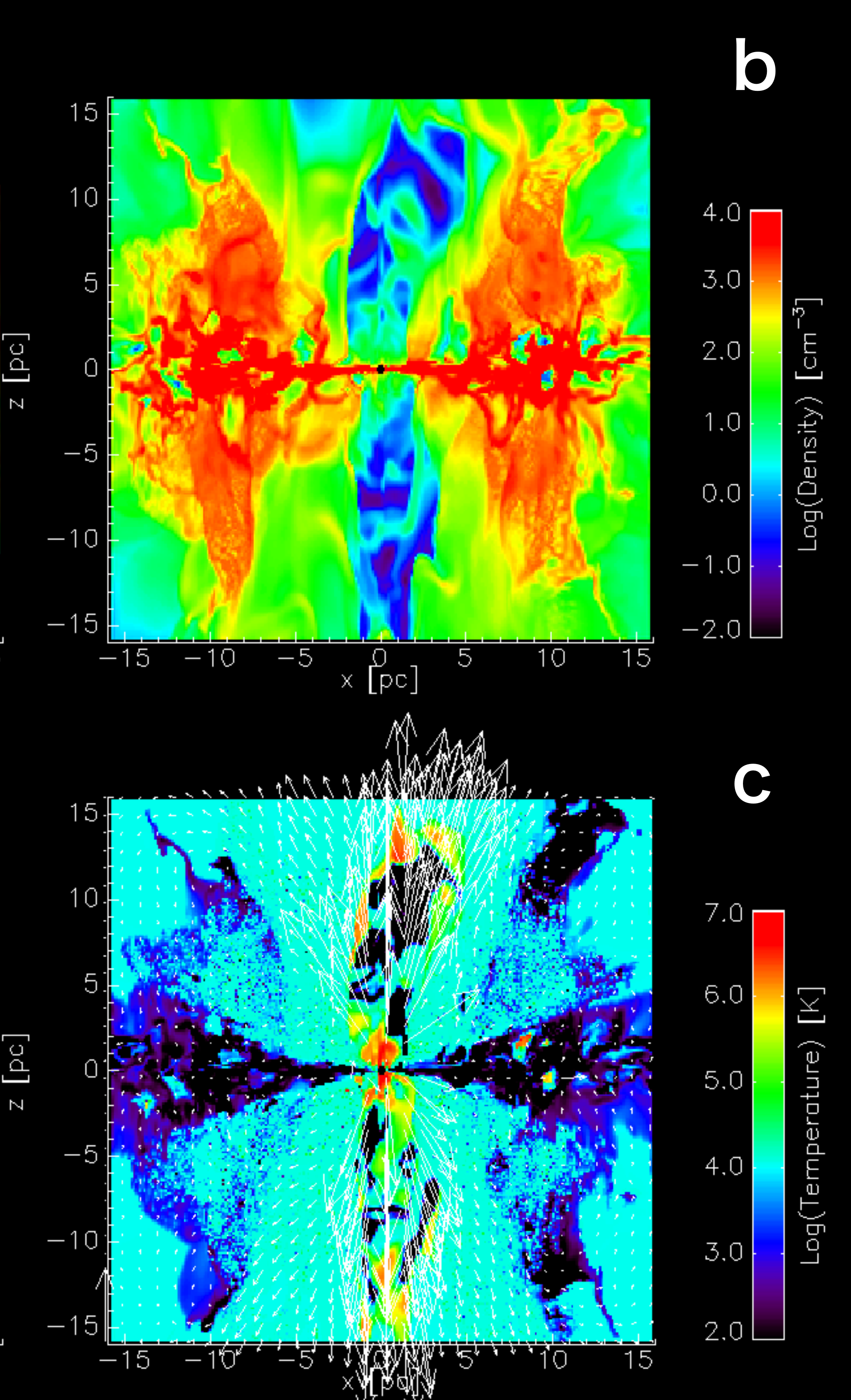}   
\caption{a) Configuration of the calculation. Radiation from the central accretion disk (denoted as AD)
 is calculated by calling C{\scriptsize LOUDY} 
along a radial ray.  (b) Three-dimensional density and (c) temperature
grid data in the $y=0$ plane is taken from a radiation-hydro simulation \citep{wada2016}. 
The rotation axis of the circumnuclear disk is along the $z$-axis.
}
\label{wada_fig: 0}
\end{figure}

%
\section{RESULTS}
%

\subsection{Intensity distributions}

\begin{figure}[h]
\centering
\includegraphics[width = 8cm]{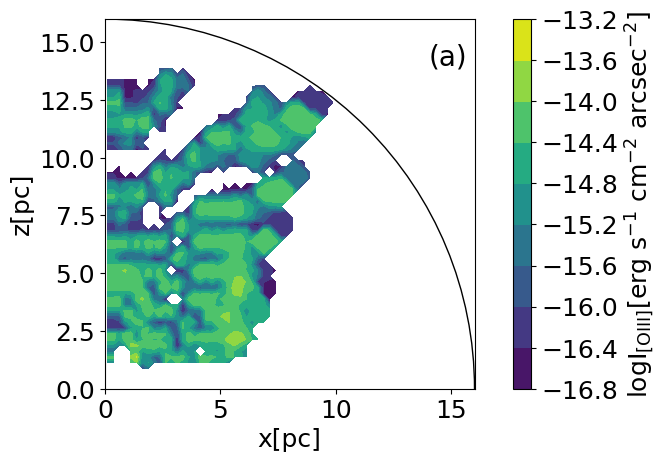} 
\includegraphics[width = 8cm]{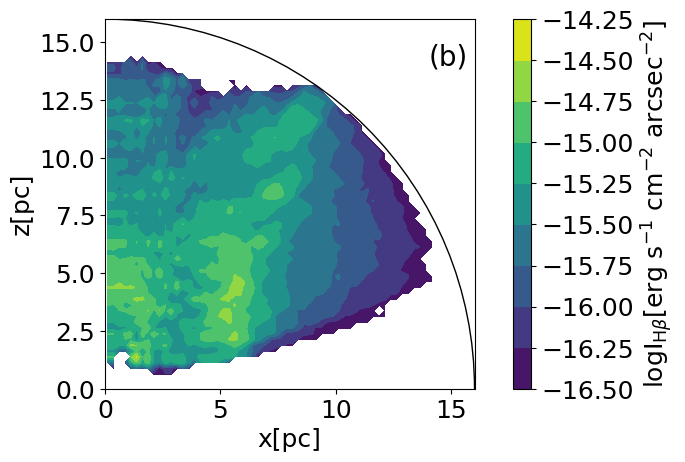} 
\includegraphics[width = 7.0cm]{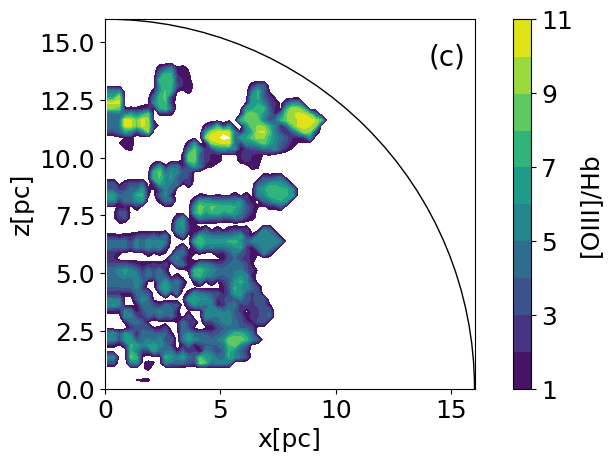} 

\caption{a) \oiii surface brightness distribution. The rotational axis points along the $z$-axis. b) Same as a), but for H$\beta$.  {c) Distribution of 
\oiiihb line ratio.}
}
\label{wada_fig: oiii}
\end{figure}

Figure \ref{wada_fig: oiii} shows the surface brightness distributions of the \oiii$\lambda 5007$ and H$\beta$ lines.
Both distributions show a conical shape around the rotation axis (i.e., $z$-axis).
The half opening angle of the \oiii emission is approximately 45 degrees.
Note that this morphology is not caused by
an optically thick ``torus", but it is 
rather a natural consequence of 
non-spherical radiation emitted from the central accretion disk
and the resulting funnel-like ‘fountain’ flows (Fig. 1b).
Radiation from a thin accretion disk and obscuration of
the surrounding material causes ionization photons to escape
into the funnel-like structures (see Fig. 1b,c and \citet{wada2012}).
In a three-dimensional view, the \oiii bright regions are distributed 
mostly at the ‘inner wall’ of the funnel flow (i.e., the yellow regions shown in Fig. 1b),
and this is also suggested from observations of near-by AGNs \citep[e.g.][]{das2006, fischer2010, crenshaw2010, muller2011}.

{
It is also notable that the internal structures of \oiii and H$\beta$
are not uniform on the scale of several parsecs, reflecting the inhomogeneous structures of 
the outflowing gas, which varies by an order of magnitude in the column density 
for a given line-of-sight in the radiation-driven fountain \citep{wada2015, wada2016}. 
Propagating the ionizing photon through the non-uniform media also 
causes variation of the spectral properties along rays (see also Fig. \ref{wada_fig: 6} and 
related discussion). 
The \oiii and H$\beta$ bright regions roughly coincide, but as 
Fig. \ref{wada_fig: oiii}c shows, 
the \oiii is relatively brighter 
with respect to the H$\beta$ at the outer part
($r \gtrsim 10$ pc)  in the ionizing cone.
}

\subsection{Ionization diagnostic diagrams}

The ionization diagnostic diagrams, or so-called ‘BPT’ diagrams \citep{baldwin1981, veilleux1987},
use two pairs of line ratios to distinguish the properties of 
emission lines in the narrow-line regions (NLRs) of Seyfert galaxies, low ionization
nuclear emission-line regions (LINERs), and starburst galaxies.
Here, we apply ionization diagnostic diagram criteria based on \citet{kewley2001} to the \oiii $\lambda 5007$, \nii$\lambda 6583$, \sii$\lambda \lambda 6717,30$, H$\alpha$, and
H$\beta$ intensities obtained by the calculations described in \S 3.1.

\begin{figure}[h]
\centering

\includegraphics[width = 8cm]{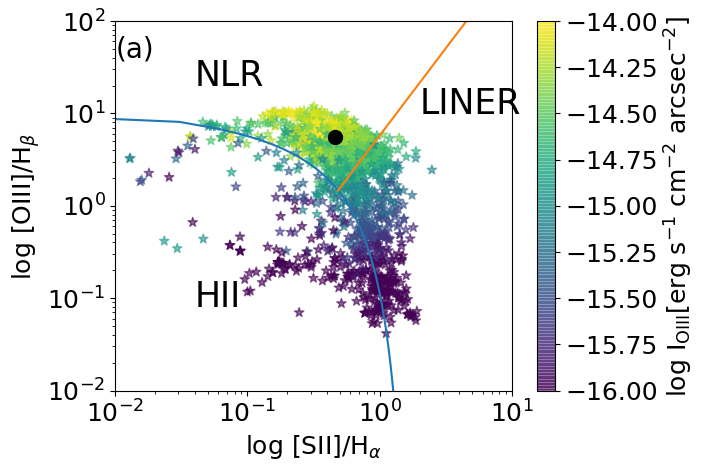} 
\includegraphics[width = 8cm]{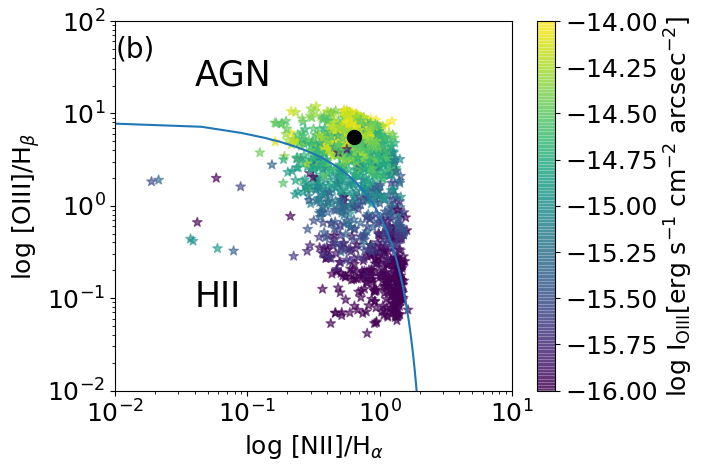} 
\includegraphics[width = 8cm]{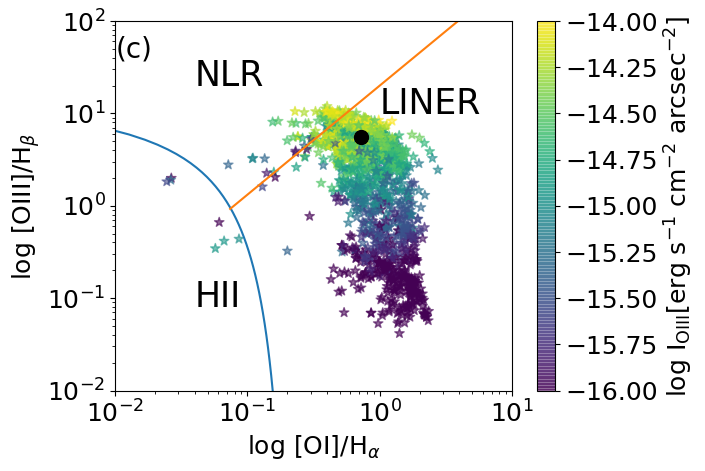} 

\caption{
{Ionization diagnostic diagrams for three emission line pairs. The solid blue curves show the starburst limit,
and the orange lines show the limit between NLRs and LINERS (\citet{kewley2001}). Color represents the \oiii surface brightness. 
The black filled circles show the \oiii brightness weighted averages.}
}

\label{wada_fig: 3}
\end{figure}

Three pairs of line ratios are plotted in Fig.\ref{wada_fig: 3}a-c. 
Note that each point represents the emission line ratios
for a given position in the observer's plane, not for {\it one galaxy}, as in the usual BPT diagrams.
These plots can be used to diagnose the {\it local} properties of the photo-ionized regions in circumnuclear ISM
on scales of a few parsecs.
In observations, similar plots have been obtained in some nearby galaxies with galactic winds using 
an integral-field spectrograph (IFU) \citep{sharp2010} (see their Fig. 21 for the Circinus galaxy),
{however, their spacial resolutions are still too large to make direct comparison with the present model (see also
Fig. \ref{wada_fig: 7})}.

{As the \oiii flux weighted average (black filled circle) shows, 
the BPT diagrams based on
\siiha (Fig. \ref{wada_fig: 3}a) and \niiha (Fig. \ref{wada_fig: 3}b)
suggest that the ionizing cone in our model can be diagnosed as 
``NLR"s or ``AGN".  However, Fig. \ref{wada_fig: 3}c shows that even the \oiii bright regions
are located in part in the ``LINER" domain (see discussion below). }

\begin{figure}[h]
\centering
\includegraphics[width = 8cm]{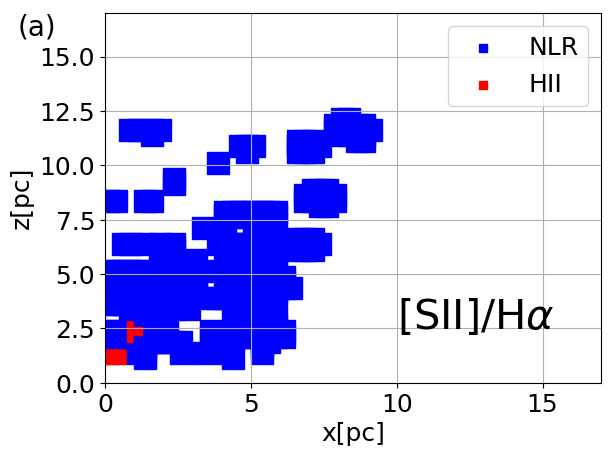}  
\includegraphics[width = 8cm]{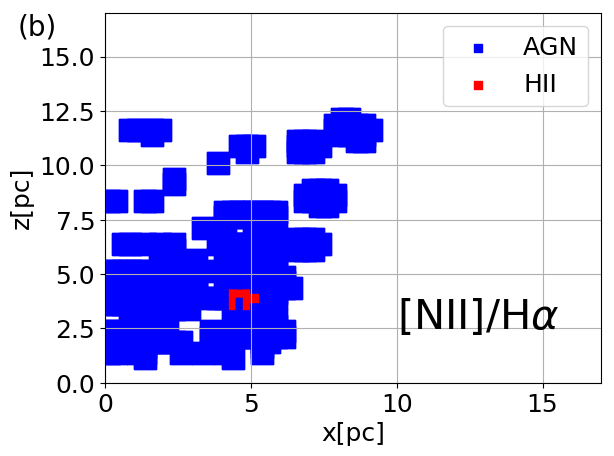}  
\includegraphics[width = 8cm]{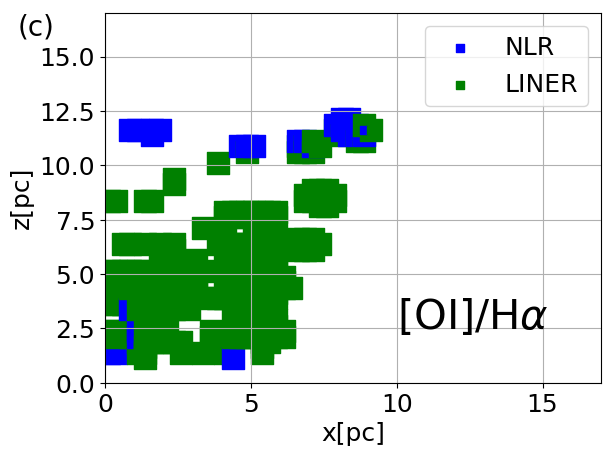} 

\caption{{a) Distribution of the three emission line regions using the \siiha- \oiiihb diagram (Fig.  \ref{wada_fig: 3}a).
b) Same as Fig.4a, but based on \niiha- \oiiihb (Fig. \ref{wada_fig: 3}b),
and c) \oiha- \oiiihb (Fig. \ref{wada_fig: 3}c).  Only \oiii bright regions  
with $ > 10^{-15}$ erg s$^{-1}$ cm$^{-2}$ arcsec$^{-2}$ are shown.}}
\label{wada_fig: 4}
\end{figure}

Figures \ref{wada_fig: 4}a-c show the distributions of the three emission line regions 
diagnosed from Figs. \ref{wada_fig: 3}a-c, respectively.
Relatively bright \oiii regions (i.e. $ > 10^{-15}$ erg s$^{-1}$ cm$^{-2}$ arcsec$^{-2}$) are plotted.
In Figs. \ref{wada_fig: 4}a and  \ref{wada_fig: 4}b , it is notable that the 
``NLR" grids (blue grids) dominate the outflow regions around the z-axis and form a conical 
shape. 
{The small patches of ``HII" (red grids) are caused by the grids with relatively less brightness in \oiii in the BPT diagram 
(Fig. \ref{wada_fig: 3}a-b). Note that this does not mean that the ionized regions in the present model 
are caused by photons from star-forming regions. In fact, we here assume only the AGN as the ionizing source, but the
incident SEDs may significantly change when the radiation radially propagates through the gas around the AGNs as 
discussed below. }

\begin{figure}[h]
\centering
\includegraphics[width = 10cm]{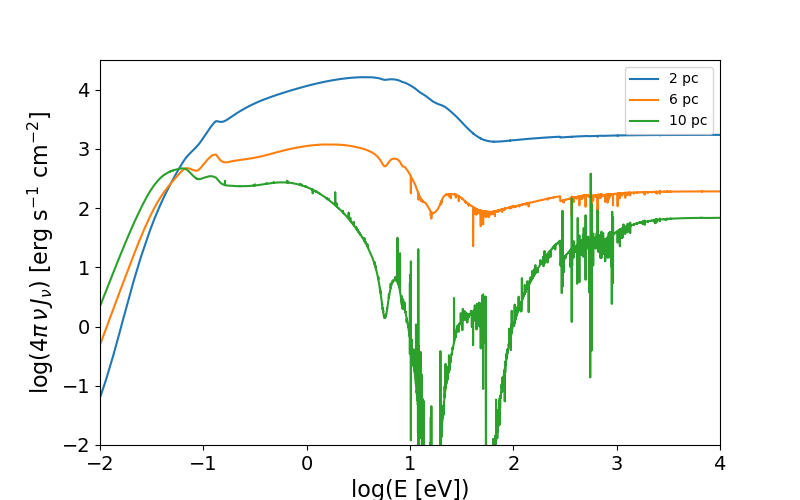}  

\caption{{Radial change of SEDs at $r= 2, 6$, and 10 pc from the center 
along a ray of $\theta = 40^\circ$ from the rotational axis.}}
\label{wada_fig: 6}
\end{figure}

{In contrast to Fig. \ref{wada_fig: 4}a and \ref{wada_fig: 4}b, the \oi -based diagram (Fig. \ref{wada_fig: 4}c) shows that
 the region at $r \lesssim 10$ pc is mostly categorized as ``LINER", reflecting
the distribution in the line-diagnostic diagram (Fig. \ref{wada_fig: 3}c), 
although some regions identified as ``NLR" are located at $r \gtrsim10 $ pc, where \oiiihb is relatively large (see Fig. \ref{wada_fig: oiii}c).
This result is seemingly paradoxical because the Circinus galaxy is not apparently LINER. 
This can be understood by the fact that \oi becomes relatively stronger relative to other lines, if UV emission
is dominated by X-ray emission in the incident SEDs \citep{nagao2002}.   
 In fact, Fig. \ref{wada_fig: 6} shows that the SEDs are heavily attenuated along a ray from the central source, especially in the UV range.
This ``filtered" radiation may change the physical conditions and line emissions of the ionized gas, depending on
the column density, as pointed out for Markarian 3 \citep{collins2009}.}

\subsection{Physical properties of the gas in NLRs}

\begin{figure}[h]
\centering
\includegraphics[width = 8cm]{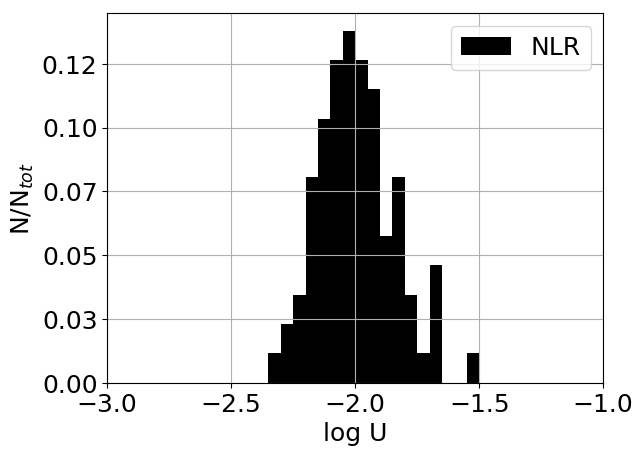} 
\includegraphics[width = 8cm]{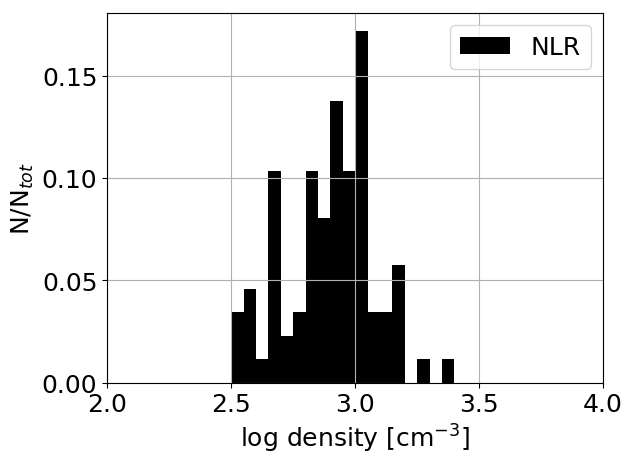}  
\includegraphics[width = 8cm]{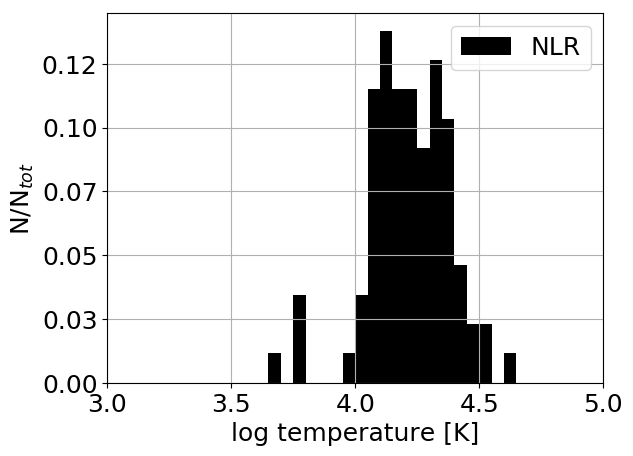} 

\caption{Histograms of the ionizing parameter $U$, density, and temperature in grid cells
categorized as NLR using
\siiha and \oiiihb.}
\label{wada_fig: 5}
\end{figure}

The physical conditions of the ISM in NLRs have been inferred from spectroscopic data in nearby AGNs.
For example, from the line ratio of \sii$\lambda\lambda 6716,6731$, the electron density 
is estimated to be $n \simeq 2000$ cm$^{-3}$\citep{koski1978}.
More recently, \citet{zhang2013} presented a statistical study based on 15,000 objects in SDSS DR7 and 
found that typical density and temperature ranges for the NLR gas are
$n \simeq 100-10^4$ cm$^{-3}$ and $T_e \simeq 1-2\times 10^4$ K, respectively. 
The origin of these line emitting gases could be AGN-driven outflows.
In the previous section, we found that emission lines from the outflow show characteristic properties of NLRs. 
The gas density, temperature, and ionizing parameter in the ``NLR" domains are 
plotted in Fig. \ref{wada_fig: 5}. 
Local density and temperature are taken from the input hydrodynamic model (i.e., 4096 grid cells).
The ionization parameter $U$, which is the ratio of the ionizing photon density to the electron density,
is defined as
\begin{eqnarray}
U \equiv \frac{1}{4\pi r^2 c\, n_{\rm H}} \int^\infty_{\nu_0} \frac{L_\nu}{h\nu} d\nu.
\end{eqnarray}
Here, $U$ is taken from each C{\scriptsize LOUDY} output.
The grid cells are counted where the \oiii local intensity is brighter than 0.3\% of its maximum value.
We found that the gas density in NLR is approximately $n \sim 300-1500$ cm$^{-3}$,
The temperature is $T \sim 1-3\times 10^4 $ K, and  $ U \sim 0.01$.
These ``NLR" gases are located mostly at the {\it inner} surface of the bicone (yellow regions in Fig 1b),
as suggested by observations of NLRs in nearby AGNs \citep[e.g.][]{das2006, muller2011}, 
{although our model here may be applied only to the innermost region 
of NLRs, extended to $\sim 100$ pc (see also the discussion in \S 4.1).}

%
\section{Discussion}
%
\subsection{Comparison with observations}
\begin{figure}[h]
\centering
\includegraphics[width = 8cm]{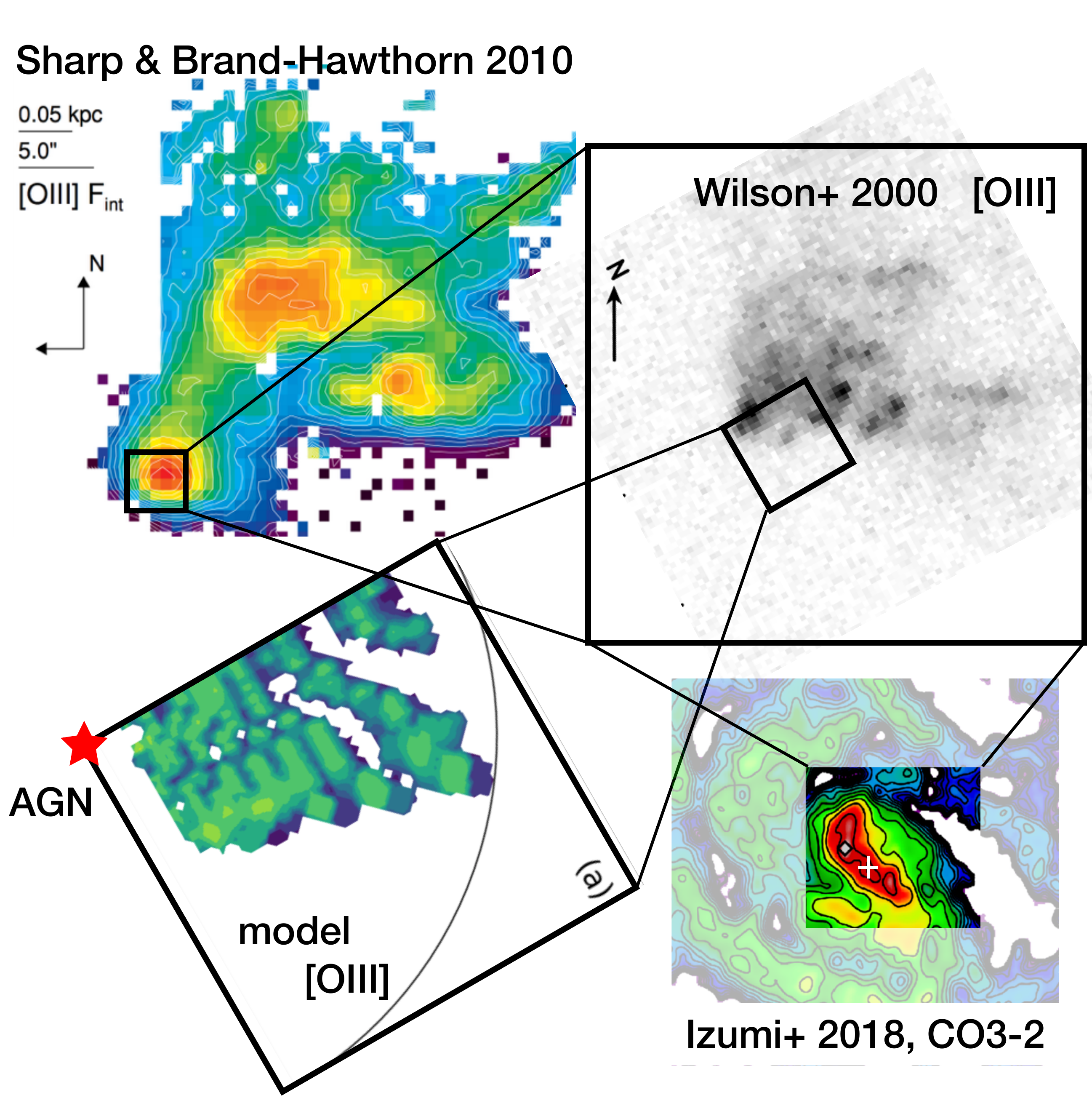} 

\caption{Comparison of Fig. \ref{wada_fig: oiii}a with observed \oiii \citep{wilson2000, sharp2010} and CO(3-2) \citep{izumi2018} in
the Circinus galaxy.}
\label{wada_fig: 7}
\end{figure}

{
In Fig. \ref{wada_fig: 7}, our \oiii map (Fig. \ref{wada_fig: oiii}a)  is compared with 
the one obtained by 
the Integral Field Unit observation using the Anglo-Australian Telescope \citep{sharp2010} and
the Hubble Space Telescope/WFPC2 \citep{wilson2000}. A CO(3-2) map obtained by ALMA \citep{izumi2018}
is also shown to elucidate the multi-phase structures in the central region of Circinus.
The IFU observations by \citet{sharp2010} show a conical ionizing region extending to $r \sim 400-500$ pc by
observing \oiii, H$\alpha$, H$\beta$, \nii, and \sii lines (see their Figs. 19 and 20).
Their observed resolution was 0.7 arcsec ($=14.3$ pc), which is ten times larger than our model.
The innermost region of the observed ionizing cone ($< 50$ pc)
is mostly categorized as ``NLR" in their ionization diagnostic diagram, but
direct comparison with Fig. \ref{wada_fig: 4} is not relevant.}
\citet{wilson2000} presented
 images of \oiii $\lambda 5007$ and  H$\alpha$ with 0.046 arcsec (0.9 pc) angular resolution in the central 20-pc region
of the Circinus galaxy using the Hubble Space Telescope/WFPC2. 
The observed \oiii ionizing cone ($< 20$ pc) has clumpy internal structures, and this is also the case in our model.
The surface brightness of the brightest spots in the \oiii cone are approximately $10^{-16}$ erg s$^{-1}$ cm$^{-2}$ pixel$^{-1}$
$\sim 10^{-14}$ erg s$^{-1}$ cm$^{-2}$ arcsec$^{-2}$,
which is comparable with that shown in Fig. \ref{wada_fig: oiii}a. 

{
The origin of the knot-like structures of the NLR seen in \citet{sharp2010} 
is still an open question. Its morphology shows some self-similarity to the
innermost structures seen in \citet{wilson2000} and also in our simulations. 
Our model presented here cannot follow the evolution of
the AGN-driven outflows on such a large scale.
Still, one should note that the outflows driven by AGNs can be intrinsically 
non-uniform and non-steady; therefore, it would be natural to see the knot-like 
structures in the emission lines. The radiative transfer effect (e.g., Fig. \ref{wada_fig: 6})
and time variation of the AGN activity itself could also enhance the
non-uniform nature in the emission lines. It is also notable that the physical conditions of 
the ``NLR" gases in our model (\S 3.3) are consistent with those that have been inferred for the observed narrow line regions.}


\subsection{Multi-phase nature in the central region}

Recently, mid-infrared (MIR) interferometric observations showed that 
 the bulk of the MIR emission comes from dust in the polar region, rather than from the dusty ``torus" \citep{tristram2014, asmus2016}.
The polar infrared emission is naturally predicted from
our radiation-fountain picture \citep{schartmann2014}. 
The present results suggest that the fountain flows can be a source of
MIR polar emission and line emission from NLRs. 
Both the ionizing cone presented here and the MIR polar emission
extend perpendicular to the molecular ``disk" observed by CO (3-2) using ALMA \citep{wada2018, izumi2018} (see also Fig. \ref{wada_fig: 7}).
The radiation-driven fountain can naturally explain the multi-phase gas structures around the nucleus, 
at least for the central region of the Circinus galaxy.
Investigating the generality of this picture for other nearby Seyfert galaxies would be an interesting subject for future theoretical and 
observational studies.

\acknowledgments
The authors would like to thank the anonymous referee for his/her constructive  
comments and suggestions.
We are grateful to T. Izumi and L. Ho for valuable comments on the manuscript.
Calculations were performed with version 17.02 of C{\scriptsize LOUDY}. We thank G. Ferland and the C{\scriptsize LOUDY} team for their continuous support. 
Numerical computations of the radiation-hydrodynamic model were performed on a Cray XC30 at the Center for Computational Astrophysics at the National Astronomical Observatory of Japan. This work was supported by JSPS KAKENHI Grant Numbers  16H03959, 18K18774, 16H01101, and 16H03958.
KW thanks Onsala Space Observatory and S. Aalto for their support and hospitality.

\end{document}